# DenseNet approach to segmentation and classification of dermatoscopic skin lesions images

Reza Zare[a,*], Arash Pourkazemi[b]

a. Faculty of Industrial and Systems Engineering, Tarbiat Modares University, Tehran, Iran.
* Corresponding author

b. Faculty of Mechanics, Electrical Power and Computer, Islamic Azad University Science and Research Branch, Tehran, Iran.

**Abstract**

At present, cancer is one of the most important health issues in the world. Because early detection and appropriate treatment in cancer are very effective in the recovery and survival of patients, image processing as a diagnostic tool can help doctors to diagnose in the first recognition of cancer. One of the most important steps in diagnosing a skin lesion is to automatically detect the border of the skin image because the accuracy of the next steps depends on it. If these subtleties are identified, they can have a great impact on the diagnosis of the disease. Therefore, there is a good opportunity to develop more accurate algorithms to analyze such images. This paper proposes an improved method for segmentation and classification for skin lesions using two architectures, the U-Net for image segmentation and the DenseNet121 for image classification which have excellent accuracy. We tested the segmentation architecture of our model on the ISIC-2018 dataset and the classification on the HAM10000 dataset. Our results show that the combination of U-Net and DenseNet121 architectures provides acceptable results in dermatoscopic image analysis compared to previous research. Another classification examined in this study is cancerous and non-cancerous samples. In this classification, cancerous and non-cancerous samples were detected in DenseNet121 network with 79.49% and 93.11% accuracy respectively.

## 1. Introduction

The skin is the body's most extensive organ, protecting the body from heat, light, and infection, helping to control its temperature, and storing fat and water. Therefore, the identification of skin diseases is of great importance. Skin cancer is one of the most common types of cancer (Berking et al, 2014; Haenssle et al., 2018). Benign tumors usually do not damage the tissue adjacent to the skin, and are generally not life-threatening. In contrast, malignant tumors can affect a person's life and must be treated because they damage nearby tissue and may even grow in other parts of the body (Linares et al., 2015). Because early detection and appropriate treatment in cancer are very effective in the recovery and survival of patients, an intelligent computer system can serve as a diagnostic tool to help doctors in the early diagnosis of cancer (Haenssle et al., 2018). Skin cancer can be found in different types including: melanoma (mel), squamous cell carcinoma (akiec), basal cell carcinoma (bcc), melanocytic nevi (nv), benign keratosis (bkl), dermatofibroma (df), and vascular lesion (vasc). Melanoma is the most unpredictable cancer that can be diagnosed at an early stage (Kaymak et al., 2018). The American Cancer Society estimates that about 101,610 people will be diagnosed with melanoma in 2021, with an estimated 7180 deaths. (Howlader et al., 2021). Melanoma is also the 19th leading cause of cancer death in the UK, accounting for 1% of all cancer deaths in the country. Accurately identifying the boundaries of skin lesions is the first essential step for computer-aided diagnostic systems (Zortea et al, 2011). The type of lesion in skin diseases has different forms that differ in size, color, symmetry, bulge and margin of the lesion. Diagnosis of the disease using artificial intelligence has different approaches, one of which is the use of a database (for example, images). This approach uses the principle of pattern recognition or deep learning in which the computer detects specific groups of clinical signs or images through iterative algorithms (Amisha et al., 2019).

Image analysis techniques allow an image to be processed so that information can be extracted automatically. These images are taken using a dermatoscope. Examples are image analysis, image segmentation, edge extraction, and texture and motion analysis (Da Silva and Mendonça, 2005). Image segmentation is difficult due to the high variability in the images (Merjulah and Chandra, 2019) and to diagnose a type of skin lesion, automatic detection of the dermatoscopic image border of the skin is one of the most important steps in diagnosis, because the accuracy of the next steps depends on it (Emre et al., 2008). The subtleties in medical images can be very complex and sometimes challenging for skilled doctors. If these subtleties are identified, they can have a great impact on the diagnosis of the disease.

Convolutional Neural Networks (CNN) have had good results for segmentation of simple images but have not progressed well for complex images. The U-Net neural network was first used to segment medical images, which, thanks to its favorable results, was later used in other fields (Ronneberger et al., 2015). Deep learning is at the forefront of artificial intelligence. With deep learning models, the level of accuracy of the algorithms will be greatly increased, so that in tasks such as image classification, the computer surpasses the human. Deep learning techniques are used in artificial neural networks. Artificial neural networks extract the patterns in the input data and determine the best result. Deep learning methods use the hidden layers of artificial neural networks. Previously, artificial neural networks had only 2 to 3 hidden layers. The number of hidden layers of deep learning now reaches 300 layers. Deep learning is part of machine learning as each layer extracts specific features and useful information from the data (Mahapatra, 2018).

The aim of this study is to present a proposed deep learning model with appropriate accuracy for image segmentation and classification of various skin diseases, which is obtained through dermatoscopic image analysis. First, we train a U-Net convolutional neural network model with appropriate accuracy using a set of skin lesion images (where the lesion boundaries are marked). Then, by generalizing this model and the best weights obtained, we segment the ground truth images of the HAM10000 data set. After dilation (in order to remove noise such as hair on the skin) and resizing the images, we classify the images using the DenseNet121 network. The research results show that the proposed model is more accurate than previous researches.

## 2. Literature review

In recent decades, the importance of diagnosing skin cancer using dermatoscopic images can be seen from research in this field. In this section, we present the research that has been done in the field of skin disease analysis using dermatoscopic images and mention the importance of their segmentation and classification, so that the participation of this research can be clearly shown.

In 2012, Sheha et al., presented two sets of dermatoscopic images for melanoma diagnosis, both automatic and manual. They first extracted the features of melanoma dermatoscopic images using the GLCM[1] and classified them manually and automatically using a multilayer perceptron neural network. In 2012, Ballerini et al., proposed an algorithm for classifying non-melanoma skin lesions based on the new hierarchical k-nearest neighbor classification. The hierarchical structure breaks down the general classification into several smaller classifications, each with specific classes. In this algorithm, feature selection is presented within the framework of hierarchical structure. This means that it extracts the characteristics related to color, texture and skin lesions from each subset. They used a set of ordinary (non-dermatoscopic) images in 5 classes to test their algorithm. In 2014, to diagnose skin cancer, Aswin et al., first improved skin dermatoscopic images (removing noise such as hair) and then sliced the images using the Threshold technique. Next, they extracted the features in dermatoscopic images using GLCM and RGB techniques and used these features to classify. They used artificial neural network classification and also used the genetic algorithm to optimize it. In 2015, Masood et al., presented a semi-supervised and self-advised model for the automatic detection of melanoma using dermatoscopic images. They used a deep belief network with labeled and unlabeled data and configured the network with an exponential error function. They used a support vector machine to classify unlabeled data. They also used SA-SVMs based on polynomials and radial bases for generalization and redundancy of the model, as well as a deep network trained with random data selected by the bootstrap method. They tested their proposed model on 100 dermatoscopic images. In 2015, Jaworek and Paweł introduced a classification system for melanocyte lesions. In addition to whether the lesion is benign or malignant, they have identified the exact type. First, they reduced the noise in the images, then segmented the skin lesions, extracted and selected the features, and finally classified. They used KNN and SVM methods for classification. They tested 300 dermatoscopic images and obtained acceptable accuracy. In 2016, Kawahara et al., showed that training the features extracted from an image using a pre-trained CNN with natural images can accurately detect a data set of 1300 images with 10 classes of skin lesions. They used ordinary (non-dermatoscopic) images and also used image normalization and data enhancement to improve accuracy. In 2018, Kaymak et al., studied the automatic diagnosis of pigmented skin lesions. Using a deep learning model, they first classified the skin lesion as melanocytic or non-melanocytic. They then identified malignant and benign types using other deep learning models. Evaluation of their performance showed a good distinction between melanocytic and non-melanocytic skin lesions. They also showed that malignant melanocytic skin lesions are classified more accurately than non-melanocytic malignant skin lesions. They used 10015 dermatoscopic images in 2-class and 3-class modes. In 2018, Rezvantalab et al., studied the effectiveness and capability of convolutional neural networks. They compared 4 different convolutional neural network architectures using 10135 images in 8 different classes. The main purpose of this study was to compare the function of deep neural network with dermatologists. These results showed that deep learning performed better than dermatologists. In another study in 2019, Brinker et al., compared the classification of skin images using a convolutional neural network with 145 dermatologists at 12 German universities. In this study, 12,378 dermatoscopic skin images in two classes were used to train the convolutional neural network with ResNet50 architecture. They used 100 clinical images to compare the function of the convolutional neural network with dermatologists. The results of the convolutional neural network with smaller variance than dermatologists showed that the computer vision is more robust compared to the human evaluation for skin image classification tasks. In 2019, Tschandl et al., in a study compared Softmax and CBIR as output layers of a convolutional neural network. In this study, they used the ResNet50 network with initial ImageNet weights. They used EDRA data set with 888 images, ISIC2017 data set with 2750 images in 3 classes and PIRV data set in 8 classes to evaluate the results. The results were similar in all three datasets based on the AUC, but for the case where the data were trained in 3 classes, the CBIR method performed better than the Softmax.

---

[1] Gray level Co-occurrence matrix

## 3. Materials and methods
### 3.1. Dataset

There are different collections of skin images that we use from two different collections in this study. We use the ISIC-2018 data set (Codella et al., 2019) to train the U-Net convolutional network. The ISIC-2018 collection consists of 2594 JPEG color dermatoscopic images. An example of these images can be seen in the fig. 1.

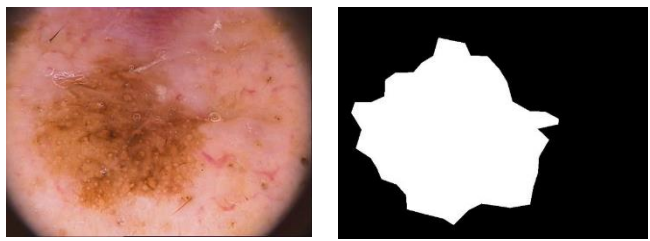

Fig. 1. An example of the images of the ISIC-2018 collection along with the corresponding ground truth

Due to the fact that the splitting of training, validation and test sets is not already available, we perform this splitting with two different ratios for this set. In the first category, we consider 70% for training data, 10% for validation data and 20% for test data. In the second category, we consider 80% of the data for training, 15% for test data and 5% for validation data (Table 1).

Table 1. Classification with two different ratios for the ISIC-2018 collection

| Category | Training | Validation | Test |
|---|---|---|---|
| 1 | 1867 | 208 | 519 |
| 2 | 2075 | 129 | 390 |

We also use the HAM10000 image collection (Tschandl et al., 2018) to classify images. The HAM10000 dataset contains 10015 dermatoscopic images collected from different populations of women (45%) and men (55%) exposed to skin diseases. Different classes of this dataset include squamous cell carcinoma, basal cell carcinoma, melanoma, benign keratosis, dermatofibroma, melanocytic nevi, and vascular lesion (Fig. 2). The table below shows the full details of the HAM10000 and its comparison with other publicly available collections.

Table 2. Compare HAM10000 image collection with other image collections

| Collection name | Number | Percentage of expert approval | Akiec | Bcc | Bkl | Df | Mel | Nv | vasc |
|---|---|---|---|---|---|---|---|---|---|
| PH2 | 200 | 20.5 | - | - | - | - | 40 | 160 | - |
| Atlas | 1024 | - | 5 | 42 | 70 | 20 | 275 | 582 | 30 |
| ISIC 2017 | 13786 | 26.3 | 2 | 33 | 575 | 7 | 1019 | 11861 | 15 |
| Rosendal | 2259 | 100 | 259 | 296 | 490 | 30 | 342 | 803 | 3 |
| ViDIR Legacy | 439 | 100 | - | 5 | 10 | 4 | 67 | 350 | 3 |
| ViDIR Current | 3363 | 77.1 | 32 | 211 | 475 | 51 | 680 | 1832 | 82 |
| ViDIR MoleMax | 3954 | 1.2 | - | 2 | 124 | 30 | 24 | 3720 | 54 |
| HAM10000 | 100015 | 53.3 | 327 | 514 | 1099 | 115 | 1113 | 6705 | 142 |

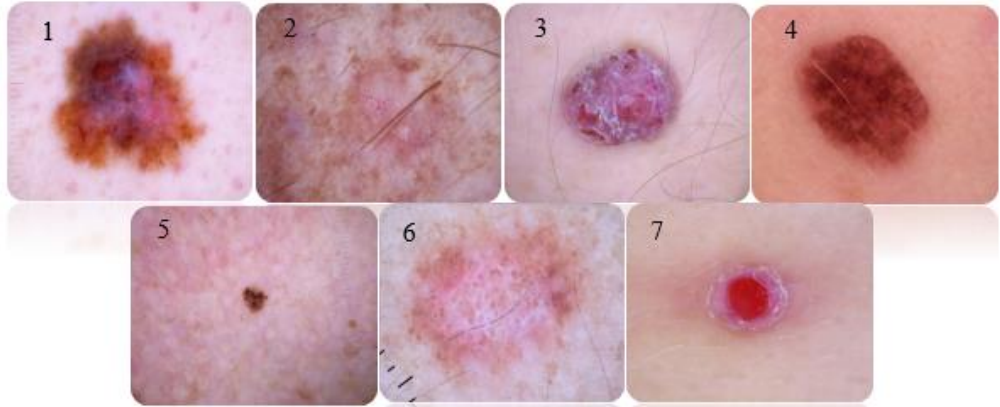

Fig. 2. Types of skin lesions [1. melanoma (mel), 2. squamous cell carcinoma (akiec), 3. basal cell carcinoma (bcc), 4. melanocytic nevi (nv), 5. benign keratosis (bkl), 6. dermatofibroma (df), 7. vascular lesion (vasc)]

### 3.2. Methodology

In this research, we use the U-Net deep learning model for image segmentation and the DenseNet121 model for image classification. The U-Net model consists of 5 contraction blocks, naturally one bottleneck and 5 expansion blocks. The final layer is a convolutional layer with one neuron and with the Sigmoid activation function for classification of 2-class, as well as another model with 7 neurons in the last layer with the softmax activation function for classification of 7-class.

#### 3.2.1. U-Net model hyper-parameters

The main parameters related to the model architecture are dropout layer rate and number of primary filters. The cost function in the proposed model is binary cross-entropy, and Adam optimizer is used. The number of epochs of this model is determined based on the progress of the model. In this way, if the accuracy of the model does not improve up to 10 epochs, the learning rate decreases by a factor of 0.01. And if the accuracy of the model still does not improve, the learning process stops. The maximum number of terms considered was 150, which went up to 48 epochs.

Table 3. U-Net model hyper-parameters

| Image size | Number of epoch | Batch size | Learning rate | Primary filters | Dropout rate |
|---|---|---|---|---|---|
| 320*224 | 150 (48) | 24 | 0.001 | 32 | 0.4 |

In this stage of convolutional neural networks (the input of this type of network is the image) which is the stage of extracting and learning the feature, first the input images of the training set pass through the network layers and by applying filters related to different layers of the network, image properties are learned hierarchically. Initially, the values of these filters, which are the same as the network weights, have random values, and after each epoch, the weights are updated according to the cost function and evaluation of the model performance on the evaluation data. In the initial or lower layers of the network, general features of the image, such as edges, are learned, and in the middle and final layers, the more detailed features of each image are learned. This process is the same as learning network weights, which can also be achieved by proper adjustment of hyper-parameters (LeCun et al., 2015).

### 3.2.2. Saving the best weights

The weights obtained from the network training, which are better than the other weights and are more accurate on the validation data during the training period, are saved. We use them to evaluate test data and transfer learning to the HAM1000 image collection. The results of the two categories mentioned earlier have been obtained in the first category with an accuracy of 94.14 and in the second category with an accuracy of 92.99.

### 3.2.3. Generalization of learning

We smooth out the gray images of the output by applying a threshold of 0.5 (more than the threshold of 255 and less than the threshold of 0) and obtain the ground truth of the images. We use the model trained by the first category because the best weights obtained from the first category are more comprehensive than the second category and the number of test data is more. Next, we examine the masks produced in the previous stage of the HAM10000 collection as a monitoring and remove the masks that are not in a single area or do not show the location of the lesion correctly from the data set. Eventually the number of images will be reduced from 10015 images to 9238 images. Details of the number of images in each class before and after the removal of noise images are given in Table 4.

Table 4. Details of the number of classes in the HAM10000 collection before and after the dropout of noise images

| Class | Number of data before deletion | Number of data after deletion | Deletion number | Deletion percentage |
|---|---|---|---|---|
| Akiec | 327 | 262 | 65 | 0.64 |
| Bcc | 514 | 388 | 126 | 1.25 |
| Bkl | 1099 | 915 | 184 | 1.83 |
| Df | 115 | 88 | 27 | 0.26 |
| Mel | 1113 | 1008 | 105 | 1.04 |
| Nv | 6705 | 6489 | 216 | 2.15 |
| Vasc | 142 | 88 | 54 | 0.53 |
| **Total** | **10015** | **9238** | **777** | **7.75** |

### 3.2.4. Data preprocessing

Apply the ground truth obtained from the previous section to the images of the HAM10000 set and cut it to the size of the lesion in the image so that the black parts of the image are removed and then resize the image to 224*224.

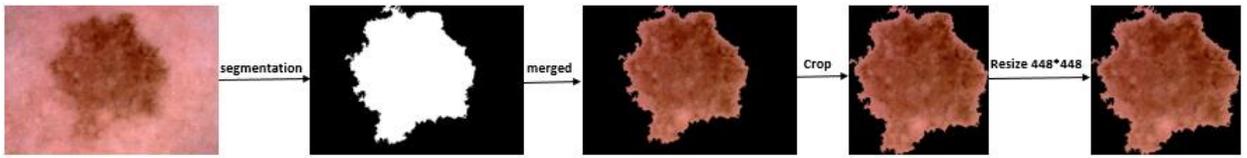

Fig. 3. Data preprocessing

Also, based on the equation (1), we finally change the image range to [-1, 1].

$$I_N = (I - MIN_I)\left(\frac{2}{MAX_I - MIN_I}\right) - 1 \qquad (1)$$

### 3.2.5. Data preparation

In this part of the research, we use the 5-fold cross validation method for the HAM10000 image collection, and for this purpose, the data in this collection are divided into two parts: training and test. Details of the number and name of classes are shown in Table 5.

Table 5. Name and number of data of each class

| Class name | Number of data in each class |
|---|---|
| Melanocytic (Mel, Nv) | 7497 |
| Non_Melanocytic (Benign, Malignant) | 1741 |
| Mel | 1008 |
| Nv | 6489 |
| Benign (df, vasc, bkl) | 1091 |
| Malignant (bcc, akiec) | 650 |
| Cancerous (Mel, Bcc, Akiec) | 1658 |
| Non-cancerous (Bkl, Df, Vasc, Nv) | 7580 |
| Akiec | 262 |
| Bcc | 388 |
| Bkl | 915 |
| Df | 88 |
| Mel | 1008 |
| Nv | 6489 |
| Vasc | 88 |

Fig. 4 shows how skin lesions are classified.

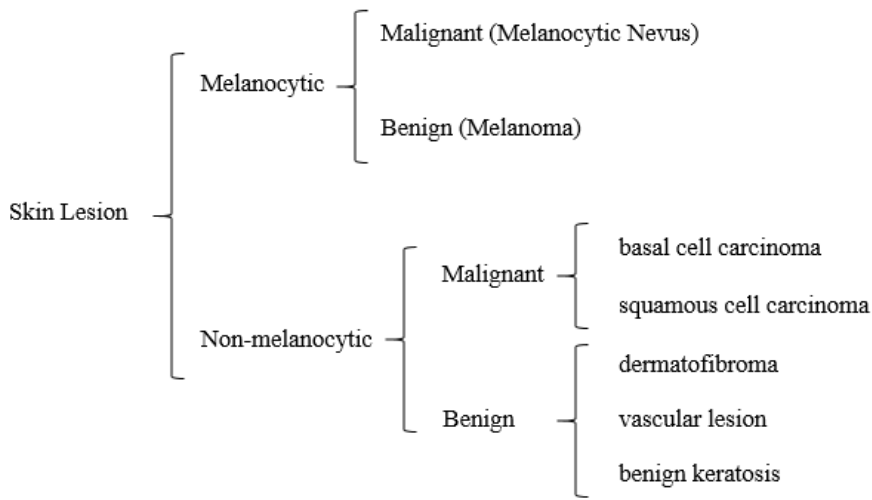

Fig. 4. ISIC-2018 challenge dataset skin lesion types (Kaymak et al., 2018)

Table 6 shows the details of the number of each fold for the 5-fold cross validation method for 2-class classification.

Table 6. Number of training and test data in each fold

| Classes | Fold size | Number of training data in each fold | Number of test data in each fold |
| --- | --- | --- | --- |
| Melanocytic/ Non_Melanocytic | 5 | 7390 | 1848 |
| Melanoma/ Melanocytic Nevus | 5 | 5998 | 1499 |
| Benign/ Malignant | 5 | 1392 | 348 |
| Cancerous/non-cancerous | 5 | 7390 | 1848 |

We divide the data into three categories, including training, testing, and validation for the 7-class mode. We consider 70% of the total data as training data, 17% as test data and 13% as validation data, which is shown in the Table 7.

Table 7. Number of data in 7-class classification

| Class | Training data (70%) | Validation data (13%) | Test data (17%) |
| --- | --- | --- | --- |
| Akiec | 183 | 32 | 47 |
| Bcc | 272 | 48 | 68 |
| Bkl | 641 | 131 | 143 |
| Df | 61 | 8 | 19 |
| Mel | 705 | 129 | 174 |
| Nv | 4542 | 840 | 1107 |
| Vasc | 61 | 8 | 19 |
| Total | 6467 | 1200 | 1571 |

## 3.3. Selection of convolutional neural network model

At this stage, the type of convolutional neural network architecture used in the training phase is determined. This research uses the latest DenseNet convolutional neural network architecture. This architecture has different versions in terms of depth and number of layers, in which the DenseNet121 version is used in this research. Modifications are also made to the architecture top layers and hyper-parameters of the model.

### 3.3.1. Hyper-parameters of DenseNet121 model

In this step, our goal is to adjust the hyper-parameters of the selective convolutional neural network to fit the data set used.

#### 3.3.1.1. The top layers of DenseNet121 for the two-class mode are as follows.

DenseNet121 → GlobalAveragePooling2D → Dense (FC) → Batch Normalization → Dropout → Dense (FC) → Sigmoid

First, a DenseNet121 model with initial ImageNet weights is used. There is a GlobalAveragePooling2D in the top layers. It is then a fully connected layer with 256 neurons and a ReLU activator. Next, we use a Batch Normalization layer and a dropout layer with a rate of 0.25. Finally, there is a fully connected layer with a neuron that perform the classification operation with the sigmoid function. Also, the RMSprop optimization algorithm with learning rates = 1e-04 and decay = 1e-06 the binary cross-entropy cost function is used.

#### 3.3.1.2. The top layers of DenseNet121 for the 7-class mode are as follows:

DenseNet121 → GlobalAveragePooling2D → Dense (FC) → Batch Normalization → Dropout → Dense (FC) → Softmax

First, a DenseNet121 model with initial imagenet weights is used. At the top layers, there is a GlobalAveragePooling2D layer. It is then a fully connected layer with 256 neurons and a ReLU activator. Then a Batch Normalization layer and a dropout layer with a rate of 0.25 are used. Finally, there is a fully connected layer with 7 neurons that perform the classification operation with the softmax function.

Table 8. Hyper-parameters of DenseNet121 model

| Classification | Image size | Number of epochs | Category size | Learning rate | Dropout rate |
|---|---|---|---|---|---|
| 2-class | 224*224 | 30 | 16 | 0.0001 | 0.25 |
| 7-class | 224*224 | 50 | 32 | 0.0006 | 0.25 |

### 3.3.2. Saving the best weights

We use the 5-fold cross validation method for two-class modes. In each category, we store the best weights obtained from the training data. Then evaluate the test data of each category. We also use the best weights for the 7-class mode to evaluate the test data.

### 3.3.3. Classification

In this step, the sigmoid function (for 2-class mode) and the softmax function (for 7-class mode) are used as activation functions in the last layer. Using the weights learned in the primary and middle layers stored from the previous step, it performs the classification operation on the test data in each category and determines the appropriate class for each input image.

### 3.3.4. Model Evaluation

In this step, we evaluate the average performance of test data using AUC, Accuracy, Sensitivity, Specificity criteria for two-class modes, as well as AUC, Precision, F1-Score and Confusion Matrix criteria for 7-class mode.

## 4. Evaluation of model

Here are the results of the proposed method for classifying different skin lesions in different conditions on the ISIC 2018 and HAM10000 image collections. In order to evaluate the proposed approaches of this research, the most widely used standard evaluation criteria in the field of disease diagnosis with the help of artificial intelligence is used. These criteria include the AUC criterion, which is used to determine the classification efficiency. Also, to show the detection rate in each class, Specificity and Specificity criteria are used for two-class modes, and Precision and F1-Score criteria are used for 7-class modes. The results of the two-class mode are presented in Table 9.

Table 9. The results of the two-class mode on the DenseNet121 network

| 2-class | Accuracy (%) | Sensitivity (%) | Specificity (%) | AUC (%) |
|---|---|---|---|---|
| Melanocytic/ Non_Melanocytic | 92.34 ± 1.76 | 95.86 ± 1.35 | 77.13 ± 7.50 | 95.69 ± 1.76 |
| Melanoma/ Melanocytic Nevus | 92.21 ± 1.10 | 73.02 ± 7.93 | 95.19 ± 2.39 | 94.58 ± 1.36 |
| Benign/ Malignant | 88.00 ± 0.99 | 92.76 ± 3.18 | 80.00 ± 7.43 | 94.52 ± 0.56 |
| Cancerous/non-cancerous | 87.70 ± 3.99 | 57.91 ± 12.70 | 94.22 ± 3.61 | 89.46 ± 7.36 |

In this section, we review the results of 7-class mode for the DenseNet121 network in normal mode. We also use micro and macro averages of Precision, F-Score and AUC criteria.

Table 10. AUC values for each class in the DenseNet121 network for 7-class mode

| Class | Normal mode |
|---|---|
| Akiec | 99.06 |
| Bcc | 99.90 |
| Bkl | 99.19 |
| Df | 99.92 |
| Mel | 97.35 |
| Nv | 98.19 |
| Vasc | 100 |
| Micro | 98.09 |
| Macro | 99.10 |

Table 11. Micro and macro values in Precision, F1-Score and AUC criteria for DenseNet121 network in 7-class mode

| Mode | Precision (%) micro – macro | F1-Score (%) micro – macro | AUC (%) micro – macro |
|---|---|---|---|
| Normal | 91.21 – 87.36 | 91.21 – 89.23 | 98.09 – 99.10 |

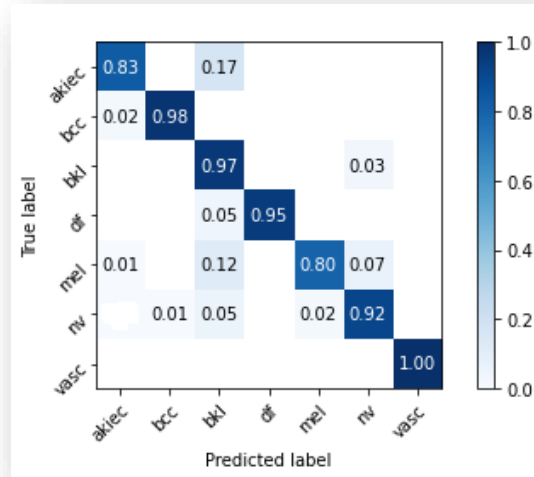

Fig. 5. Confusion Matrix in DenseNet121 network for 7_class mode in normal mode

### 4.1. Comparison with Different Frameworks

In this step, the results obtained in three binary-classification modes are compared with the results obtained from the study of Kaymak et al., (2018).

Table 12. Comparison of the proposed method with the research of Kaymak et al. (2018)

| 2-class | Reference | Number of Train | Number of Test | Accuracy % | Sensitivity % | Specificity % |
|---|---|---|---|---|---|---|
| Melanocytic/ Non_Melanocytic | Kaymak et al, 2018 | 13099 | 2004 | 78 | 83.9 | 63.5 |
|  | DenseNet121 (5_Fold) | 7390 | 1848 | 93.02 ± 0.96 | 95.20 ± 2.43 | 83.97 ± 5.96 |
| Melanoma/ Menocytic Nevus | Kaymak et al, 2018 | 11854 | 1564 | 84 | 84.7 | 83.8 |
|  | DenseNet121 (5_Fold) | 5998 | 1499 | 93.36 ± 0.54 | 76.68 ± 6.82 | 95.95 ± 1.29 |
| Benign/ Malignant | Kaymak et al, 2018 | 1757 | 440 | 58 | 60.6 | 57.8 |
|  | DenseNet121 (5_Fold) | 1392 | 348 | 89.26 ± 0.97 | 91.66 ± 1.85 | 85.23 ± 2.85 |

For example, in the method of Kaymak et al., (2018), the accuracy of the diagnoses is 58% and the values of Sensitivity and Specificity were 60.6% and 57.8% respectively, while in the proposed approach, the results are much better.

Table 13. Comparison of the number of training and test samples in the proposed method with Kaymak et al. (2018) research in Melanoma and Menocytic Nevus classes

| Reference | Number of Train for Mel | Number of Train for Nv | Number of Test for Mel | Number of Test for Nv |
|---|---|---|---|---|
| Kaymak et al, 2018 | 6400 | 5454 | 313 | 1251 |
| DenseNet121 (5_Fold) | 807 | 5191 | 201 | 1298 |

Kaymak et al. (2018) increased the number of melanoma training samples to 6400 images using data augmentation and the number of Nv training samples was 5454, but in our proposed method, the number of melanoma samples in each category is equal to 807 images. Table 13 shows the details of the number of training and test samples in each class.

In this section, the results of the proposed approach are compared with the research of Rezvantalab et al., (2018) in the 7-class mode. In their research, 10,215 images (HAM10000 and PH2 image collections) and DenseNet201 network were used. As can be seen in Table 14, the results of the proposed method are better than the results of the research of Rezvanatalab et al., (2018).

Table 14. Comparison of the proposed method with the research of Rezvanatalab et al., (2018) in 7-class mode

| Reference | Precision (%) Micro - Macro | F1-Score (%) Micro - Macro | AUC (%) Micro - Macro |
|---|---|---|---|
| Rezvan Talab et al (2018) – DenseNet201 | 89.01 – 85.24 | 89.01 – 85.13 | 98.79 – 98.16 |
| DenseNet121 | 94.02 – 89.39 | 94.02 – 91.59 | 98.96 – 99.26 |

Another classification examined in this study is cancerous and non-cancerous samples. There are three cancer classes and four non-cancer classes in 7 classes of the HAM10000 image collection (Table 5). In this classification, cancerous samples were detected in DenseNet121 network with 79.49% accuracy and non-cancerous samples with 93.11% accuracy.

## 5. Conclusions

Accurate identification of skin lesion boundaries is the first essential step for computer-aided diagnostic systems. This paper surveyed the ability of deep learning to classify skin lesions. In this paper, we developed an improved algorithm for segmentation and classification for skin lesions using CNN. They are used in dermatoscopic images and have excellent accuracy. We tested the segmentation architecture of our model on the ISIC-2018 dataset and the classification on the HAM10000 dataset. Our results show that the combination of U-Net and DenseNet121 architecture provides acceptable results in dermatoscopic image analysis.

# References

Amisha, P. M., Pathania, M., & Rathaur, V. K. (2019). Overview of artificial intelligence in medicine. *Journal of family medicine and primary care*, *8*(7), 2328.

Aswin, R. B., Jaleel, J. A., & Salim, S. (2014, July). Hybrid genetic algorithm—Artificial neural network classifier for skin cancer detection. In *2014 International Conference on Control, Instrumentation, Communication and Computational Technologies (ICCICCT) (pp. 1304-1309)*. IEEE.

Ballerini, L., Fisher, R. B., Aldridge, B., & Rees, J. (2012, May). Non-melanoma skin lesion classification using colour image data in a hierarchical K-NN classifier. In *2012 9th IEEE International Symposium on Biomedical Imaging (ISBI)* (pp. 358-361). IEEE.

Berking, C., Hauschild, A., Kölbl, O., Mast, G., & Gutzmer, R. (2014). Basal cell carcinoma—treatments for the commonest skin cancer. *Deutsches Ärzteblatt International*, *111*(22), 389.

Brinker, T. J., Hekler, A., Enk, A. H., Klode, J., Hauschild, A., Berking, C. & Schrüfer, P. (2019). A convolutional neural network trained with dermoscopic images performed on par with 145 dermatologists in a clinical melanoma image classification task. *European Journal of Cancer*, *111*, 148-154.

Codella, N., Rotemberg, V., Tschandl, P., Celebi, M. E., Dusza, S., Gutman, D., & Halpern, A. (2019). Skin lesion analysis toward melanoma detection 2018: A challenge hosted by the international skin imaging collaboration (isic). *arXiv preprint arXiv:1902.03368*.

Eduardo A.B. da Silva, Gelson V. Mendonça. The electrical engineering handbook. CRC press, 2005.

Emre Celebi, M., Kingravi, H. A., Iyatomi, H., Alp Aslandogan, Y., Stoecker, W. V., Moss, R. H., ... & Menzies, S. W. (2008). Border detection in dermoscopy images using statistical region merging. *Skin Research and Technology*, *14*(3), 347-353.

Haenssle, H. A., Fink, C., Schneiderbauer, R., Toberer, F., Buhl, T., Blum, A., & Zalaudek, I. (2018). Man against machine: diagnostic performance of a deep learning convolutional neural network for dermoscopic melanoma recognition in comparison to 58 dermatologists. *Annals of oncology*, *29*(8), 1836-1842.

Jaworek-Korjakowska, J., & Kłeczek, P. (2016). Automatic classification of specific melanocytic lesions using artificial intelligence. *BioMed research international*, *2016*.

Kawahara, J., BenTaieb, A., & Hamarneh, G. (2016, April). Deep features to classify skin lesions. In *2016 IEEE 13th international symposium on biomedical imaging (ISBI)* (pp. 1397-1400). IEEE.

Kaymak, S., Esmaili, P., & Serener, A. (2018, November). Deep learning for two-step classification of malignant pigmented skin lesions. In *2018 14th Symposium on Neural Networks and Applications (NEUREL)* (pp. 1-6). IEEE.

LeCun, Y., Bengio, Y., & Hinton, G. (2015). Deep learning. *Nature 521 (7553), 436-444*.

Linares, M. A., Zakaria, A., & Nizran, P. (2015). Skin cancer. *Primary care*, *42*(4), 645-659.

Mahapatra, S. (2018). Why deep learning over traditional machine learning. *Towards Data Science*.

Masood, A., Al-Jumaily, A., & Anam, K. (2015, April). Self-supervised learning model for skin cancer diagnosis. In *2015 7th International IEEE/EMBS Conference on Neural Engineering (NER)* (pp. 1012-1015). IEEE.

Merjulah, R., & Chandra, J. (2019). Classification of myocardial ischemia in delayed contrast enhancement using machine learning. In *Intelligent Data Analysis for Biomedical Applications* (pp. 209-235). Academic Press.

Rezvantalab, A., Safigholi, H., & Karimijeshni, S. (2018). Dermatologist level dermoscopy skin cancer classification using different deep learning convolutional neural networks algorithms. *arXiv preprint arXiv:1810.10348*.

Ronneberger, O., Fischer, P., & Brox, T. (2015, October). U-net: Convolutional networks for biomedical image segmentation. In *International Conference on Medical image computing and computer-assisted intervention* (pp. 234-241). Springer, Cham.

Sheha, M. A., Mabrouk, M. S., & Sharawy, A. (2012). Automatic detection of melanoma skin cancer using texture analysis. *International Journal of Computer Applications*, *42*(20), 22-26.

Tschandl, P., Rosendahl, C., & Kittler, H. (2018). The HAM10000 dataset, a large collection of multi-source dermatoscopic images of common pigmented skin lesions. *Scientific data*, *5*(1), 1-9.

Zortea, M., Skrøvseth, S. O., Schopf, T. R., Kirchesch, H. M., & Godtliebsen, F. (2011). Automatic segmentation of dermoscopic images by iterative classification. *International Journal of Biomedical Imaging*, *2011*.